Letter

# Interference with Symbol-misalignment

Shengli Zhang[†‡], Soung-Chang Liew[†], Lu Lu[†], and Hui Wang[‡]

[†] The Department of Information Engineering, the Chinese University of Hong Kong, Hong Kong

[‡] Communication Engineering Department, Shenzhen University, China

**Abstract:**

This paper studies the impact of interference asynchrony among different links in a wireless network. Without deliberate coordination and cooperation among the active links, there is a naturally occurring misalignment between the symbols of the targeted signal of a receiver and the symbols of the interfering signals. Interestingly, we show that the interference asynchrony can actually improve the BER performance, compared with the situation in which symbols of al signals ay aligned. In particular, we show that symbol misalignment can decrease the "effective interference power" and change the distribution of the interfering signals, in a way that results in lower BER. To ensure that symbol misalignment can be consistently attained, we propose two simple schemes that introduce time-varying symbol offsets to obtain an "average" performance of random symbol misalignment. Notably, our schemes do not change the simple receiver design structure; only the transmitters are modified in a minor way.

# I. Introduction

This paper studies the impact of asynchronous interference among different links in a wireless network. By interference asynchrony, we refer to the situation in which the symbols from different transmitters are not aligned at a receiver, so that there is a relative offset between the symbols of the targeted signal and the symbols of interfering signals. This phenomenon occurs naturally in a simple wireless network where the links do not cooperate to align their symbols. Interestingly, we show that the interference asynchrony actually helps to improve the BER performance. To fully exploit this advantage, we show how interference asynchrony can be introduced by simple modification of the transmitter design, without modification of the receiver design.

At a first glance, interference asynchrony might appear to be harmful to the system. Indeed, in CDMA networks, the orthogonality between different spreading sequences may be decreased when the boundaries of the spread symbols from different sources are not aligned. In this paper, however, we are interested in non-CDMA networks in which the signature waveforms used by different transmitters are the same. This is the case, for example, in a wireless local network. Our goal is to study and resolve "collisions" in such networks when more than one transmitter transmits. In this case, the time offset between the interference and the target signal effectively introduces a new semi-orthogonal signature wave, making it easier for the receiver to extract the desired signal.

We note that asynchrony between multiple signals have been studied and exploited under the framework of multi-user detection (MUD) [1-4]. This paper differs from the asynchronous MUD scheme in that each receiver only decodes its target signal and never tries to decode the interfering signals. The complexity of our scheme is comparable to that of a simple point-to-point scheme in which no additional information about the interfering signals is needed. By contrast, the receiver in MUD also decodes the interfering signals, and consequently its decoding structure is more complex and requires additional information on the interfering signals (e.g., pulse shaping function, spread code, and channel code). There have also been investigations on co-channel interference [5, 6] that focus on the number of interferences. Unlike our paper here, [5, 6] did not study the dependency of the system performance on a given value of symbol misalignment --- only the system performance averaged over all misalignment has been studied. Furthermore, [5, 6] did not discuss how this "average" performance can be achieved.

The rest of this paper is organized as follows. Section II formulates and analyzes the effect of symbol misalignment. We show that both the "effective" power of the interference and the distribution of the interference are modified by interference asynchrony, in a way that results in improved BER performance. However, in a simple wireless networks in which the transmitters do not cooperate, we cannot count on the availability of sufficient interference asynchrony to improve system performance. To solve this problem, we propose in Section III two simple schemes that introduce time-varying symbol offsets to obtain an "average" performance of random symbol misalignment. Our schemes do not change the simple receiver design structure; only the transmitters are modified in a minor way. Section IV concludes this paper.

## II. System Model and Characterization of Interference - Symbol Misalignment

This section first introduces the system model under study. After that, we propose two new metrics, "effective interference power" and "effective SINR", to better characterize the effect of interference under symbol misalignment. We then derive the distribution of interference under a given symbol offset and analyze the associated BER performance. Finally, numerical results are given to verify our analysis.

*A. System Model*

To convey the concept we attempt to put forth in the simplest manner, we focus on a two-link network as shown in Fig. 1. In the figure, transmitting nodes $T_1$ and $T_2$ simultaneously send information to their respective receiving nodes, $R_1$ and $R_2$. Thus, in this scenario $T_1$ is the interfering node of $R_2$, and $T_2$ is the interfering node of $R_1$. Besides its simplicity, this model is also of interest in practice. For cellular networks, it was shown that under an interference-limited (as opposed to noise-limited) environment, the interference is typically dominated by a single interfering node [7]. For IEEE 802.11 networks, the probability of *m*-transmitter collisions ($m \geq 3$) is much less than the probability of 2-transmitter collisions, thanks to the exponential backoff mechanism that dynamically adjusts the transmission probability according to the busyness of the network.

We assume that both transmitters send a sequence of *N* independent symbols with BPSK modulation at unit transmit power. The symbol duration is *T*. We assume time-invariant flat fading channels. The channel coefficients of both links, $T_1$-$R_1$ and $T_2$-$R_2$, are normalized to 1. The channel coefficient between $T_1$ and $R_2$ is denoted by a real number $h_2$, and the channel coefficient between $T_2$ and $R_1$ is denoted by another real number $h_1$, and we assume $h_1 < 1$, $h_2 < 1$. Note that, the results in this paper still hold under QPSK modulation with complex channel coefficients. Given the symmetry between link 1 and link 2, the discussion in this paper will focus only on link 1.

According to the system model in Fig. 1, the overall base-band signal received by $R_1$ can be expressed as

$$y(t) = \sum_{n=0}^{N-1} \left[ a_{1,n} f_1(t-nT) + h_1 a_{2,n} f_2(t-nT-\tau) \right] + w(t) \quad (1)$$

where $f_i(t)$ is the time-invariant pulse shaping transmit filter of node $T_i$, $a_{i,n}$ is the *n*-th BPSK modulated symbol of $T_i$, $\tau \in [0,T)$ is the symbol offset of the interfering signal relative to the information signal, and *w(t)* is the additive white Gaussian noise. For simplicity, both $f_1(t)$ and $f_2(t)$ are set to rectangle pulse shaping filter with unit energy, i.e.,

$$f_1(t) = f_2(t) = \begin{cases} 1/T & \text{if } 0 \leq t < T \\ 0 & \text{else} \end{cases} \quad (2)$$

We assume a simple receiver design that does not make use of MUD. We will show that even with a conventional receiver, a non-zero symbol offset $\tau$ will to our advantage. In a conventional non-MUD receiver design, $R_1$ tries to synchronize to its desired signal $\sum_{n=0}^{N-1} a_{1,n} f_1(t-nT)$, and then

filters $y(t)$ with a matched filter and samples it every $T$ period. The resulting discrete signal is

$$y(n) = \begin{cases} a_{1,n} + h_1 \dfrac{T-\tau}{T} a_{2,n} + w(n) & \text{when } n = 0 \\ a_{1,n} + h_1 \left( \dfrac{\tau}{T} a_{2,n-1} + \dfrac{T-\tau}{T} a_{2,n} \right) + w(n) & \text{when } n \geq 1 \end{cases} \qquad (3)$$

where the sampled noise $w(n)$ is Gaussian distributed with mean zero and variance $\sigma^2$.

## B. Effective SINR (Interference Power) and Interference Distribution

SINR, defined as the ratio of signal power to interference plus noise power, is an important metric for system performance. Traditionally, the power of the interference is calculated by averaging power symbol by symbol, in a way that is independent of the target signal. For example, the power of the interference for $R_1$ is $h_1^2$ (the transmit power of $T_2$ is unit according to our system model). Accordingly, the SINR is defined conventionally as

$$SINR = \frac{1}{h_1^2 + \sigma^2}. \qquad (4)$$

Both the power of interference and the SINR are independent to the relative symbol offset $\tau$.

*Interference power and effective SINR:*

A closer look of (3) indicates that the power of the "effective interference" to the target signal is

$$P_I = \mathrm{E}\left\{ h_1 \left( \frac{\tau}{T} a_{2,n-1} + \frac{T-\tau}{T} a_{2,n} \right) \right\}^2 = h_1^2 (2\delta^2 - 2\delta + 1) \qquad (5)$$

where $\delta = \dfrac{\tau}{T} \in [0,1)$. Note that in (5) we have assumed large $N$ so that the effect of the first symbol can be ignored. Obviously, $P_I$ not only depends on the channel coefficient $h_1$, but also on the symbol misalignment $\tau = \delta T$. It is easy to verify that $P_I$ is always less than $h_1^2$, and it is symmetric about $\delta = 0.5$. Accordingly, we refer to the SINR computed from (5) as the effective SINR (*eSINR*), given as follows:

$$eSINR(\delta) = \frac{1}{P_I + \sigma^2} = \frac{1}{h_1^2(2\delta^2 - 2\delta + 1) + \sigma^2} \qquad (6).$$

From (6), we can see that when the transmit power of the interfering node and the Gaussian noise are fixed, the SINR still depends on the symbol misalignment, $\delta$, between the interfering signal and the target signal. The SINR is maximized when $\delta = 0.5$, and minimized when $\delta = 0$ or $1$. Note that, the effective interference power and the effective SINR do not depend on the modulation schemes, i.e., the results in (5) and (6) hold for all constellation maps, not just BPSK.

*Interference distribution:*

As shown in [8], besides SINR, the system BER performance may also be affected by the distributions of noise and interference. The symbol misalignment not only changes the interference power, but also the distribution of interference. Without symbol misalignment, the interference is either $h_1$ or $-h_1$, with equal probability; with symbol misalignment, the interference is $h_1$, $-h_1$, $h_1(1-2\delta)$, or $-h_1(1-2\delta)$, with equal probability. Besides the reduction of effective interference power, the change in the distribution of interference due to symbol misalignment can also affect the BER performance.

*C. BER Analysis*

In Part B, we provided the power and distribution of interference as functions of symbol misalignment in closed form. Here, we analyze the uncoded BER performance (which is of direct interest in a communication system) and show that symbol misalignment has a significant impact on it.

For BPSK demodulation, we use the simple sign decision as in a traditional receiver (this is also the optimal maximum likelihood demodulation method). Then, the $n$-th estimated bit at $R_1$ is

$$\hat{a}_{1,n} = sign(y(n)) = \begin{cases} sign\left(a_{1,n} + h_1(1-\delta)a_{2,n} + w(n)\right) & n=0 \\ sign\left(a_{1,n} + h_1(\delta a_{2,n-1} + (1-\delta)a_{2,n}) + w(n)\right) & n \geq 1 \end{cases} \quad (7)$$

For the first symbol (i.e., $n=0$), the error probability is

$$p_{e1} = \Pr(a_{1,0} = -1)\Pr(h_1 a_{2,0} + w(0) \geq 1) + \Pr(a_{1,0} = 1)\Pr(h_1 a_{2,0} + w(0) \leq -1)$$
$$= \Pr(h_1 a_{2,0} + w(0) \geq 1)$$
$$= \Pr(a_{2,0} = 1)\Pr(w(0) \geq 1 - h_1) + \Pr(a_{2,0} = -1)\Pr(w(0) \geq 1 + h_1) \quad (8)$$
$$= \frac{1}{2}\left[Q\left(\frac{1-h_1(1-\delta)}{\sigma}\right) + Q\left(\frac{1+h_1(1-\delta)}{\sigma}\right)\right]$$

For the other symbols (i.e., $n \geq 1$), the error probability is

$$\Pr(a_{1,n} = -1)\Pr(h_1 \delta a_{2,n-1} + h_1(1-\delta)a_{2,n} + w(n) \geq 1) + \Pr(a_{1,n} = 1)\Pr(h_1 \delta a_{2,n-1} + h_1(1-\delta)a_{2,n} + w(n) \leq -1)$$
$$= \Pr(h_1 \delta a_{2,n-1} + h_1(1-\delta)a_{2,n} + w(n) \geq 1)$$
$$= \Pr(a_{2,n} = 1)\Pr(a_{2,n-1} = 1)\Pr(w(n) \geq 1 - h_1) + \Pr(a_{2,n} = -1)\Pr(a_{2,n-1} = -1)\Pr(w(n) \geq 1 + h_1)$$
$$+ \Pr(a_{2,n} = 1)\Pr(a_{2,n-1} = -1)\Pr(w(n) \geq 1 - h_1(1-2\delta)) + \Pr(a_{2,n} = -1)\Pr(a_{2,n-1} = -1)\Pr(w(n) \geq 1 + h_1(1-2\delta))$$
$$= \frac{1}{4}\left[Q\left(\frac{1-h_1}{\sigma}\right) + Q\left(\frac{1+h_1}{\sigma}\right) + Q\left(\frac{1-h_1(1-2\delta)}{\sigma}\right) + Q\left(\frac{1+h_1(1-2\delta)}{\sigma}\right)\right]$$
$$(9)$$

Combining (8) and (9), we can obtain the closed form BER expression as a function of $\delta$

$$P_e(\delta) = \frac{1}{2N}\left[Q\left(\frac{1-h_1(1-\delta)}{\sigma}\right) + Q\left(\frac{1+h_1(1-\delta)}{\sigma}\right)\right] + \frac{N-1}{4N}\left[Q\left(\frac{1-h_1}{\sigma}\right) + Q\left(\frac{1+h_1}{\sigma}\right) + Q\left(\frac{1-h_1(1-2\delta)}{\sigma}\right) + Q\left(\frac{1+h_1(1-2\delta)}{\sigma}\right)\right] \quad (10)$$

When $N$ is large, as in practical systems, $P_e$ above approaches the result in (9).

From (9), we can see that the uncoded BER performance depends on the symbol misalignment $\delta$ when the transmit power of the interfering node and the power of the Gaussian noise are fixed. As with *eSINR*, $P_e(\delta)$ is symmetric about $\delta = 0.5$. Since $Q(x)$ is a convex function, we can easily prove that the error probability in (9) is minimized when $\delta = 0.5$, and maximized when $\delta$ approaches 0 or $T$, as shown in Fig. 2.

*D. Numerical Simulation*

We now present numerical simulation results to illustrate the effect of $\delta$ on the BER performance of the two-link system. In Fig. 2, we show the BER performance under different settings of SIR ($1/|h_1|^2$) and symbol misalignment ($\delta$) when the SNR ($1/\sigma^2$) is fixed to 10 dB. In Fig. 3, we show the BER performance under different settings of SNR and symbol misalignment

($\delta$) when the SIR is fixed to 4dB. From both figures, we can see that the simulation results (the circles in the figure) and the theoretical results (the lines in the figure) match very well. As the symbol misalignment $\delta$ increases from 0 to 0.5, the BER performance improves significantly. For any given BER in Fig. 2, the SIR improvement between $\delta=0$ and $\delta=0.5$ increases from about 2dB to 2.5dB as the SIR increases. In Fig. 3, the SNR improvement decrease from about 3.5dB to 1.5dB as the power of noise decreases.

## III. Performance Enhancement by Randomization of Symbol Misalignment

In Section II, we see that the symbol misalignment between the target signal and the interfering signal can significantly improve the BER performance. However, in simple wireless network designs, we should not assume there is collaboration between the receiver and the interfering node to tune for optimal $\delta$. Without collaboration, we can only assume that $\delta$ is uniformly distributed from 0 to 1. Since $\delta$ is the same for all symbols in one block (packet), the situation is somewhat like that of a block fading channel [9] (one of the most deleterious channels) in that once a bad $\delta$ is encountered, the unfortunate happenstance persists through the whole block.

For a block fading channel, the transmitter has to transmit at a rate low enough so that the packet can be correctly decoded with high probability at the receiver. Similarly, for our random misalignment, the transmitter must adjust its data rate (or its transmit power) by assuming the worst symbol misalignment, i.e., $\delta = 0$, to guarantee the successful reception of all the packets.

*A. Two Schemes for Removing Deleterious Effect of Blockwise Symbol Misalingment*

We now propose two simple schemes to get rid of the aforementioned deleterious effect of blockwise symbol misalignment based on the system model in Fig. 1 (they can be easily extended to general WLAN). The main idea in our schemes is for a transmitter to gradually extend its symbol duration in one packet. This diversifies the values of $\delta$ within a block so that effectively all possible $\delta$ values between 0 and 1 are experienced (by different symbols). As a result, an average performance can just be achieved[1], and our system design will not be at the mercy of the worst-case $\delta$.

*Scheme A:*

With reference to Fig. 1, we assume $T_1$ transmits in the conventional way, and $T_2$ extends its each symbol duration by $\alpha = T/N$ for each successive symbol. We assume that the parameter $\alpha$ is only known to $T_2$'s partner $R_2$, and no communication between the two links is needed. Receiver $R_1$ works in the same way as a traditional receiver, and receiver $R_2$ works similarly except that it extends the matched filtering time by $k\alpha$ for the $k$-th symbol during its receiving process. Thus, at each receiver, the symbol misalignment $\delta$ varies over values from zero to 1 with step size $\alpha$.

*Scheme B:*

*Scheme A* implies off-line coordination between the two links to decide which link should extend its symbol duration. To remove this requirement, we propose another scheme which employs a random extension mechanism. Consider an integer $K$, $0 \ll K \ll N$, known by both transmitters. Before transmission, each link $i$ randomly selects an integer $K_i$ from [0, $K$]. Then the transmitter $T_i$ extends its symbol duration by $K_i \alpha$ for each successive symbol, in a similar way as $T_2$ does in *Scheme A*; $R_i$ also works accordingly. In this scheme, the symbol misalignment $\delta$ in one packet varies among the values from zero to $|K_1 - K_2|T$ with step size $|K_1 - K_2|\alpha$.

*B. Performance Analysis*

This part analyzes the SINR and BER performance of the two schemes. We assume both $K$ and $N/K$ are very large and consider the asymptotic performance as they approach infinity. Then the links in the two schemes should have similar performance since each receiver experiences all possible interference misalignments and achieves the average performance. Take link 1 in *Scheme A* for example. Assuming the initial symbol misalignment is $\delta_0$, the interfering-symbol misalignment at the $n$-th symbol of $T_1$'s packet can be expressed as

$$\delta(n) = \begin{cases} \delta_0 + n/N & \text{if } \delta_0 + n/N < 1 \\ \delta_0 + n/N - 1 & \text{otherwise} \end{cases} \quad (11)$$

---

[1] Without channel coding, an average uncoded BER can be achieved; with channel coding, an average capacity can be achieved.

Eq. (12) shows that the symbols within the packet of link 1 would experience almost all possible symbol misalignment when $1/N$ is small enough. As a result, all possible interference power and interference distribution caused by the symbol misalignment is experienced by the packet. In other words, with our proposed schemes, the block-fading like interference is changed to ergodic fast-fading like interference. With $N$ approaching infinity, the averaged *eSINR* within one packet is

$$\overline{eSINR} = \int_0^1 eSINR(\delta) d\delta$$
$$= \int_0^1 \frac{1}{h_1^2(2\delta^2 - 2\delta + 1) + \sigma^2} d\delta = \sqrt{\frac{4}{(h_1^2 + 2\sigma^2)h_1^2}} \arctan \sqrt{\frac{h_1^2}{h_1^2 + 2\sigma^2}} \quad (12)$$

The averaged uncoded BER can be expressed as

$$\overline{P_e} = \int_0^1 P_e(\delta) d\delta =$$
$$\frac{1}{4} Q\left(\frac{1-h_1}{\sigma}\right) + \frac{1}{4} Q\left(\frac{1+h_1}{\sigma}\right) + \frac{1}{4} \int_0^1 Q\left(\frac{1-h_1(1-2\delta)}{\sigma}\right) + Q\left(\frac{1+h_1(1-2\delta)}{\sigma}\right) d\delta \quad (13)$$
$$= \frac{1}{4} Q\left(\frac{1-h_1}{\sigma}\right) + \frac{1}{4} Q\left(\frac{1+h_1}{\sigma}\right) + \frac{1}{2} \int_0^1 Q\left(\frac{1-h_1(1-2\delta)}{\sigma}\right) d\delta$$

*C. Numerical Simulation*

In Fig. 4, we show the BER performance of the proposed schemes and compare it with a setting without symbol misalignment. In both figures, we find that the analytical results in (13) (the lines in the figures) match the simulation results (the circles in the figures) very well. For any given BER in Fig. 4, the proposed schemes have an SNR improvement of at least 1.5 dB.

In Fig. 5, we show the analytical (the lines) and simulated (the circles) *eSINR* (power ratio in (6)) for the proposed schemes when the SIR changes from 5 to 15 dB with the SNR ($1/\sigma^2$) is fixed to 10 dB. For the purpose of comparison, the *eSINR* of the worst case with $\delta = 0$ is also given. From this figure, we can see that the SINR improvement of the proposed schemes is more than

1.5dB.

## IV. Conclusion

This paper shows that when the transmission of a link is interfered by the transmission of another link in a wireless network, symbol misalignment between the two signals can improve BER performance. When the misalignment is varied from 0 symbol to 0.5 symbol, the "effective" interference power decreases, increasing the likelihood of correctly detecting the targeted signal.

We derive closed forms of SINR and BER as functions of symbol misalignment. In addition, we propose two new transmission schemes to randomize the symbol misalignment within a packet so that we can consistently achieve the "average" BER performance. Our schemes do not require coordination and cooperation between the links. Compared with the scenario without symbol misalignment, about 1.5dB SIR improvement can be achieved by our schemes when the SNR is 10dB.

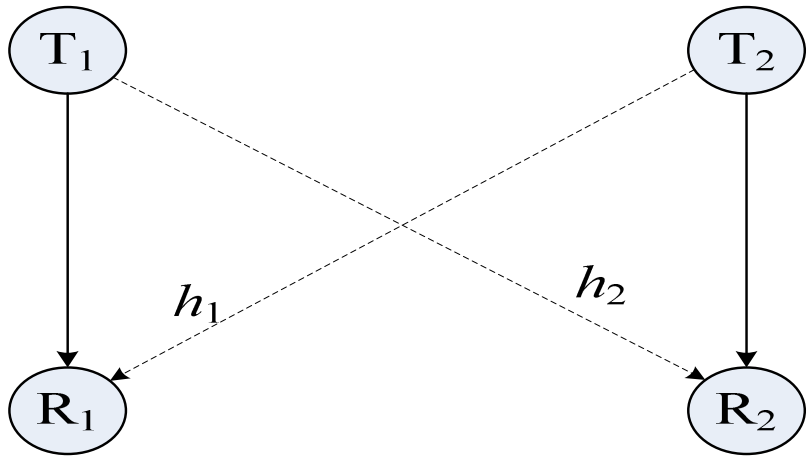

Fig. 1, Two-link interference system

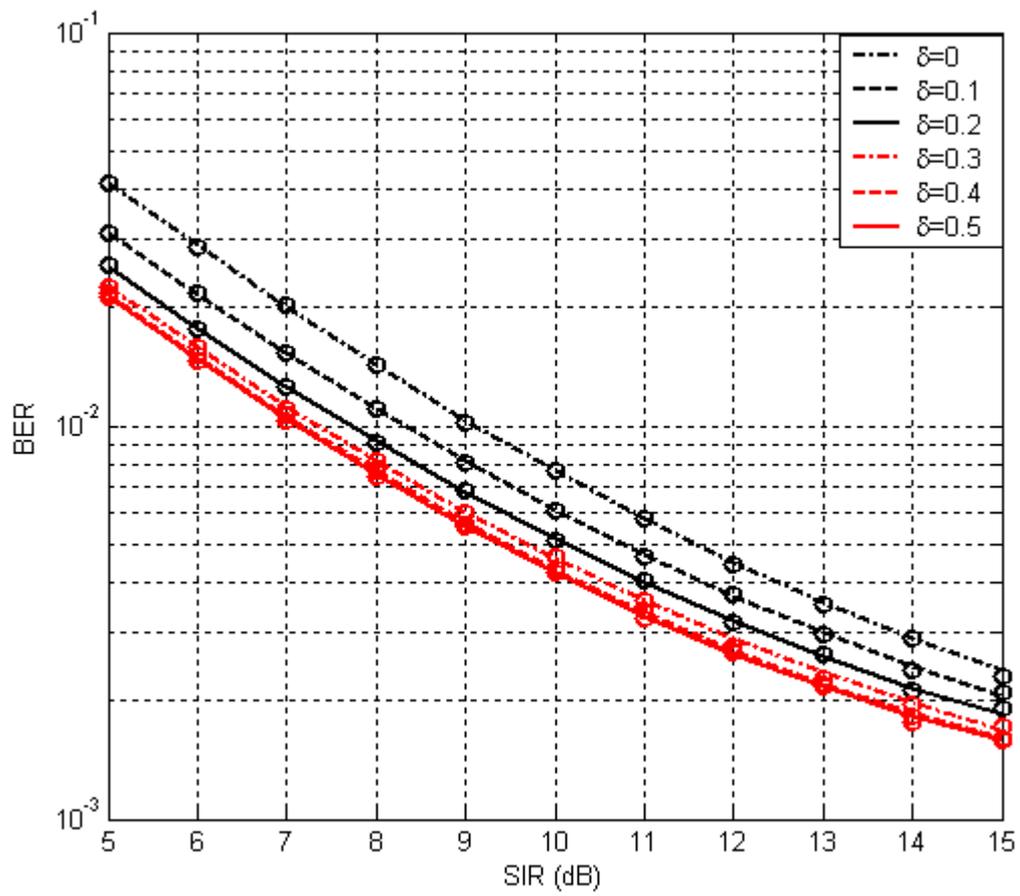

Fig. 2　BER versus SIR when SNR is 10 dB. The circles in the figure are simulation results and the lines are theoretical results.

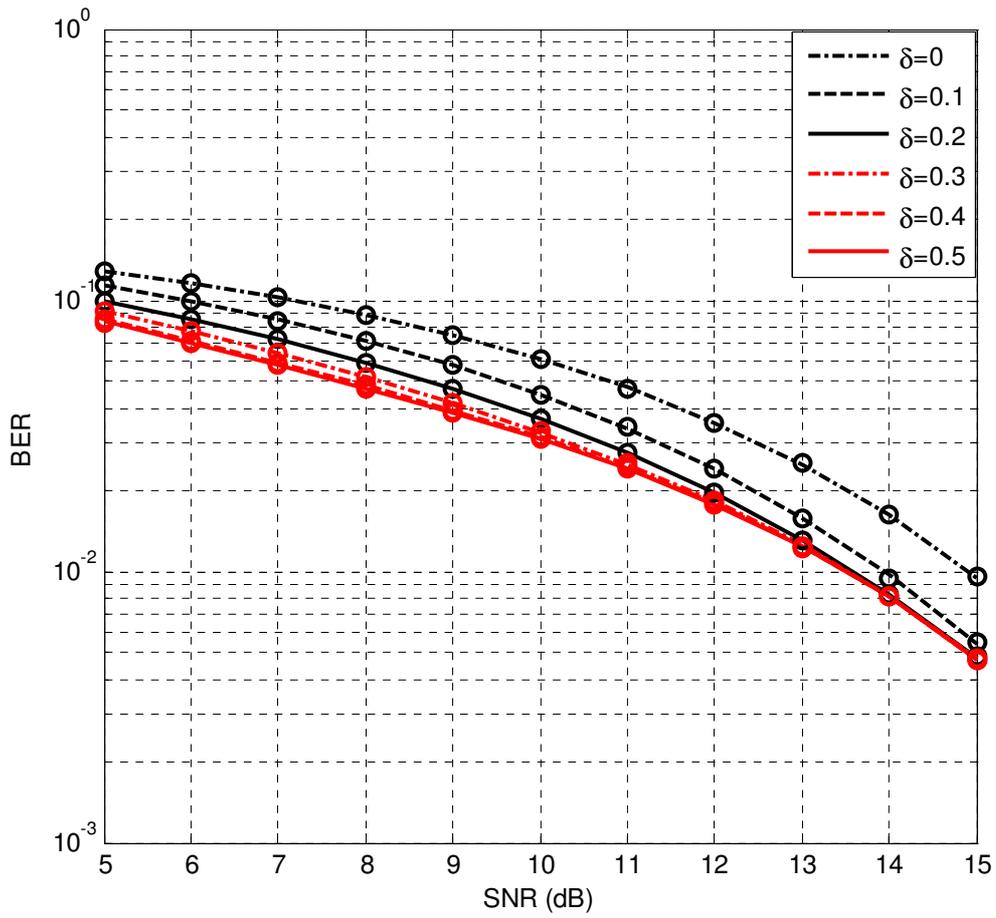

Fig. 3　BER versus SNR when SIR is 4dB. The circles in the figure are simulation results and the lines are theoretical results.

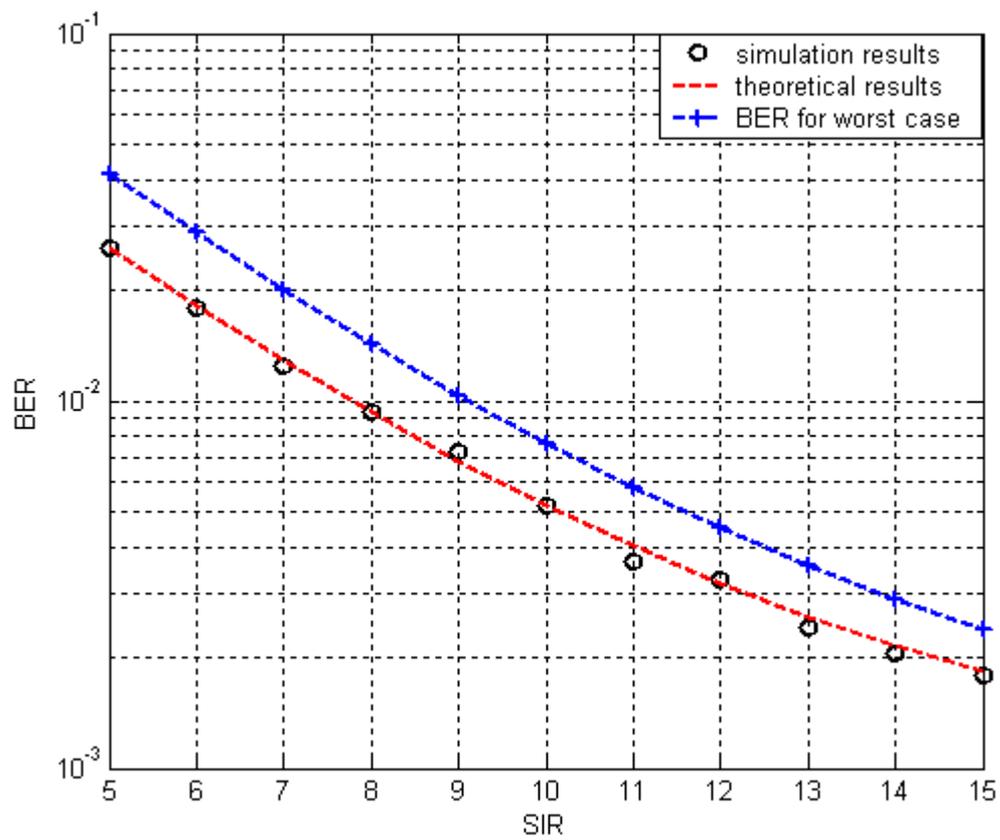

Fig. 4. BER performance of our proposed symbol-misalignment randomization schemes. SNR is fixed to 10 dB. The red line is the theoretical results, and the circles along the red line are simulation results. The blue line is the scenario when there is no symbol misalignment ($\delta = 0$).

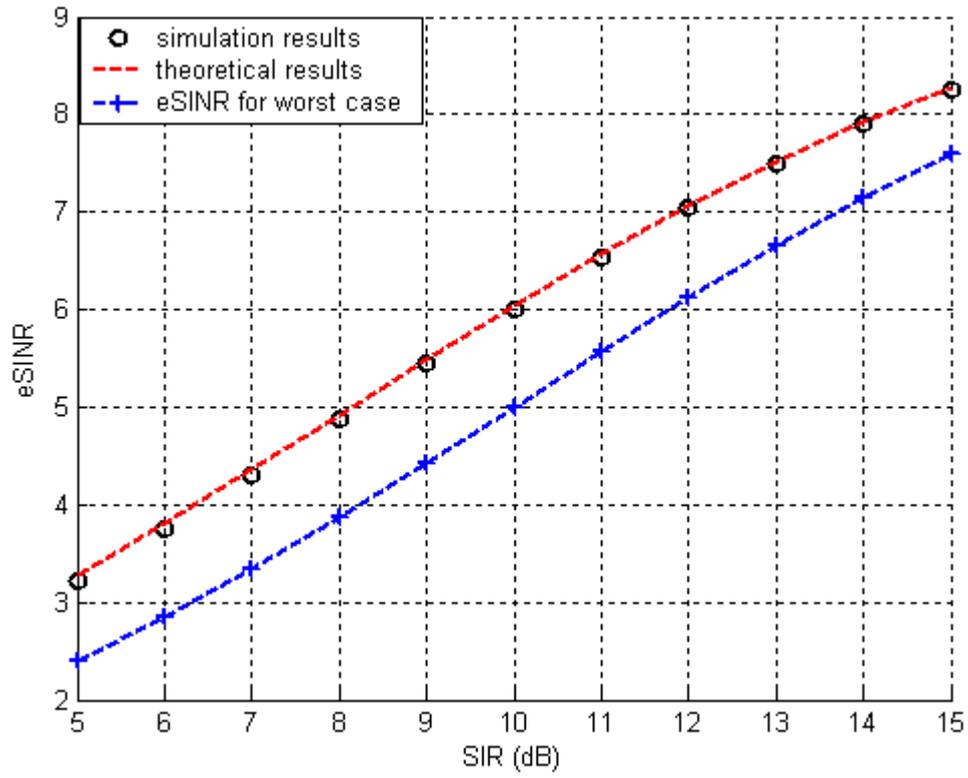

Fig. 5. eSINR of our proposed symbol-misalignment randomization schemes. SNR is fixed to 10dB. The red line is the theoretical results, and the circles along the red line are simulation results. The blue line is the scenario when there is no symbol misalignment ($\delta = 0$).